\documentclass[journal]{IEEEtran}

\usepackage{graphicx}
\usepackage{float}
\usepackage{mathrsfs}
\usepackage{amsfonts}
\usepackage{color}
\usepackage{multirow}
\usepackage{cite}
\usepackage{caption}
\usepackage{graphicx} %use graph format
\usepackage{epstopdf}

\usepackage{amsmath}
\usepackage{subfigure}

\ifCLASSINFOpdf

\else

\fi

\begin{document}

\title{Game-Theoretical Analysis of Mining Strategy for Bitcoin-NG Blockchain Protocol}

\author{\IEEEauthorblockN{Taotao Wang, \IEEEmembership{Member, IEEE}, Xiaoqian Bai, Hao Wang, \IEEEmembership{Member, IEEE},
Soung Chang Liew, \IEEEmembership{Fellow, IEEE}, Shengli Zhang, \IEEEmembership{Senior Member, IEEE}}% <-this % stops a space

%\thanks{This work was supported in part by the National Natural Science Foundation of China under grants 61701311 and 61771315, in part by the Guangdong NSF Project under grant 2016KZDXM006. (\em{Corresponding Author: Shengli Zhang})}

\thanks{T. Wang, X. Bai and S. Zhang are with the Guangdong Laboratory of Artificial Intelligence and Digital Economy (SZ), Shenzhen University, Shenzhen 518060, China (e-mail: ttwang@szu.edu.cn; baixiaoqian\_szu@163.com; zsl@szu.edu.cn).
	
H. Wang is with the Department of Data Science and Artificial Intelligence, Faculty of Information Technology, Monash University, Melbourne, VIC 3800, Australia (e-mail: Hao.Wang2@monash.edu).
	 
S. Liew is with the Department of Information Engineering, The Chinese University of Hong Kong, Hong Kong SAR, China (e-mail: soung@ie.cuhk.edu.hk)}% <-this % stops a space

}

% make the title area
\maketitle

\begin{abstract}
Bitcoin-NG (Next Generation), a scalable blockchain protocol, divides each block into a key block and many micro blocks to effectively improve the transaction processing capacity. Bitcoin-NG has a special incentive mechanism (i.e. splitting transaction fees to the current and the next leader) to maintain its security. However, this incentive mechanism ignores the joint effect of transaction fees, mint coins and mining duration lengths on the expected mining reward. In this paper, we identify the advanced mining attack that deliberately ignores micro blocks to enlarge the mining-duration length to increase the likelihood of winning the mining race. We first show that an advanced mining attacker can maximize its expected reward by optimizing its mining-duration length. We then formulate a game-theoretical model in which multiple mining players perform advanced mining to compete with each other. We analyze the Nash equilibrium for the mining game. Our analytical and simulation results indicate that all mining players in the mining game converge to having advanced mining at the equilibrium and have no incentives for deviating from the equilibrium; the transaction processing capability of Bitcoin-NG at the equilibrium is decreased by advanced mining. Therefore, we conclude that the Bitcoin-NG blockchain protocol is vulnerable to advanced mining.
\end{abstract}


\IEEEpeerreviewmaketitle

\section{Introduction}
% no \IEEEPARstart
\IEEEPARstart{B}{ITCOIN} \cite{nakamoto2008bitcoin} the first successful decentralized digital cryptocurrency, has gained much recognition and support from people in various fields. It has become the 11th largest currency in the world, with a market capitalization of over 0.21 trillion US dollars as of August 2019. As the foundation technology for Bitcoin, blockchain is a decentralized and distributed digital ledger that stores data in chronological order in a way that the data in the chain cannot be falsified. Blockchain has become a cutting-edge technology in the fields of FinTech \cite{fanning2016blockchain}, Internet of Things (IoT) \cite{ferrag2018blockchain, dai2019blockchain}, and supply chains \cite{abeyratne2016blockchain}, thanks to its ability to enable Byzantine agreement over a permissionless decentralized network \cite{garay2015bitcoin}.

Despite its strong security and privacy protection, Bitcoin blockchain faces a significant scalability problem, i.e., the speed at which it can handle transactions is restricted by the block size and block interval \cite{decker2013information, sompolinsky2013accelerating, bagaria2018deconstructing}. In Bitcoin blockchain network, miners devote computational powers to solve a hash puzzle in each round. The miner who has successfully solved the hash puzzle becomes the leader for that round and broadcasts a block that contains transactions to the whole network. For the current Bitcoin blockchain protocol, the maximum size of each block is set to 2 MB and the average interval between two successive blocks is fixed to 10 minutes, which means that Bitcoin blockchain can only handle up to 8 Transactions Per Second (TPS), given a typical transaction size of 250 Bytes. This TPS is a very low transaction processing capacity, compared to the average $2000$ TPS of Visa global payment system. 

In order to improve its on-chain transaction processing capacity, Bitcoin blockchain could simply increase the block size or reduce the block interval. However, increasing the block size (by packing more transactions into each block) and reducing the block interval (by decreasing the difficulty of hash puzzles) both lead to more forks on blockchain, which compromises the security of blockchain. Without redesigning the blockchain protocol, it is hard to increase the transaction processing capacity of Bitcoin blockchain by simply tuning these protocol parameters. 

To solve the scalability problem of Bitcoin blockchain, many new blockchain protocols have been proposed. For the detail discussions about the existing blockchain protocols, we refer the interested reader to the surveys \cite{consensus2019, natoli2019deconstructing} and the references therein. Among these blockchain protocols, Bitcoin-NG (Next Generation) \cite{eyal2016bitcoin} blockchain has attracted much attention, thanks to its effectiveness in solving the blockchain scalability problem and its compatibility with the current Bitcoin blockchain protocol \cite{eyal2016bitcoin}. To achieve a large transaction processing capacity, Bitcoin-NG decouples each block into two types of blocks: a key block and a number of micro blocks. The key block is used to elect the leader for this round. The micro blocks are used to record transactions onto the blockchain. According to the Bitcoin-NG blockchain protocol, the first miner that correctly solves the current hash puzzle can create a new key block and becomes the leader for the current round. After placing the new key block on top of the previous block, the leader is in charge of packing transactions into the following micro blocks. The creation of micro blocks does not require the mining process of solving hash puzzles. Bitcoin-NG blockchain can achieve very fast transaction processing speed, since transactions are packaged into micro blocks that are released much faster than key blocks.

The incentive mechanism of Bitcoin-NG is different from that of Bitcoin in which all transaction fees are allocated to the leader of the current round. In Bitcoin-NG blockchain, the incentive mechanism distributes a part of the transaction fees contained in the micro blocks to the leader of the current round and the remaining part of the transaction fees to the leader of the next round \cite{eyal2016bitcoin}. With this special incentive mechanism, Bitcoin-NG blockchain encourages miners to behavior honestly, i.e., to follow the default behaviors of extending the heaviest chain, including transactions, and extending the longest chain \cite{eyal2016bitcoin}.
The Bitcoin-NG blockchain protocol can significantly improve the transaction processing capacity of blockchain networks. However, we point out that Bitcoin-NG blockchain is vulnerable to advanced mining attack that compromises its security. Advanced mining attack refers to the mining behavior in which miners ignore some micro blocks issued by the leader of the current round and intentionally mine the next key block in advance to enlarge their lengths of mining duration (see details in Section IV). Although mining in advance will earn less transaction fees (since it ignores and discards some micro blocks), it will increase the probability of mining success and thus increases the rewards from the mint coins contained in key blocks. Therefore, there is still a motivation for miners to perform advanced mining attack. Without considering the joint effect of mining-duration length, mint coins contained in  key blocks, transaction fees contained in micro blocks, the original design of Bitcoin-NG blockchain \cite{eyal2016bitcoin} is not robust against advanced mining attack.

In this paper, we conduct a thoughtful analysis of the advanced mining behavior for the Bitcoin-NG blockchain protocol, by taking all the relevant factors (i.e., transaction fees, mint coins, mining time lengths) into account. Specifically, we have the following three contributions.

\begin{itemize}
  \item First, we analyze the scenario where an attacker can mine the next key block in advance, while other miners in the network follow honest mining. We use this scenario to explain what is the advanced mining attack and why it is more profitable. We formulate the attacker's mining problem as an optimization problem that aims to maximize the expected mining reward with respect to the length of mining duration. We find the optimal length of mining duration to get the maximum expected reward. Our results show that the attacker can indeed gain more rewards under the optimal advanced mining strategy than honest miners do. 
  \item Second, we proceed to analyze the scenario where the whole Bitcoin-NG blockchain network is divided into two mining pools, and both mining pools can adopt advanced mining. The analysis gets different as each pool's revenue is affected by the mining-duration length of the other. We then formulate the mining process of the two-pool scenario as a two-player game. We analytically find the Nash equilibrium for this two-player mining game. We further extend the two-player game to an $N$-player game that models the general scenario of $N>2$ mining pools. We find that there always exist a Nash equilibrium for this $N$-player game, although its analytical result is hard to derive. 
  \item Third, we perform numerical computations and system simulations to investigate and verify our analytical results. For different computing power profiles, we numerically compute the optimal mining lengths and corresponding maximum expected rewards of the attacker's mining optimization problem; we numerically compute the equilibrium points of the mining game. We also construct a Bitcoin-NG simulator to investigate the advanced mining problem. We simulated $2^{10}$ miners mining at identical rates that are divided into two and three mining pools who perform advanced mining to compete with each other. Our numerical and simulation results confirm our analytical results and suggest how to alleviate the negative effect of advanced mining for the Bitcoin-NG blockchain protocol.
\end{itemize}
The remainder of this paper is organized as follows. Section II gives a blockchain preliminary. Section III reviews the Bitcoin-NG blockchain protocol. Section IV presents our analysis for the advanced mining problem. Section V provides numerical and simulation results and Section IV concludes this paper.

\section{Blockchain Preliminary}
Blockchain is first proposed as the decentralized append-only ledger for the crypto-currency, Bitcoin. The data of blockchain is replicated and shared among all participants. Its past recorded data are tamper-resistant and participants can only append new data to the tail-end of the chain of blocks. The state of blockchain is changed according to transactions issued by the payers. Specifically, the issued transactions are broadcasted over the blockchain network. Participants then collect and group these transactions into blocks and append them to the blockchain. Each block contains a header and a set of transactions. The header of the block encapsulates the hash of the preceding block, the hash of this block, the merkle root of all transactions contained in this block, and a number called nonce that is generated according to the consensus protocol of Proof-of-Work (PoW) \cite{nakamoto2008bitcoin}. Since each block must refer to its preceding block by placing the hash of its preceding block in its header, the sequence of blocks then forms a chain arranged in a chronological order. Fig. 1 illustrates the data structure of Bitcoin blockchain.

\subsection{Proof of Work and Mining}
Bitcoin blockchain adopts the PoW consensus protocol to validate new blocks in a decentralized manner. In each round, the PoW protocol selects a leader that is responsible for packing transactions into a block and appends this block to blockchain. To prevent adversaries from monopolizing blockchain, the leader selection must be approximately random. Since blockchain is permissionless and anonymity is inherently designed as a goal of blockchain, it must consider the sybil attack where an adversary simply creates many participants with different identities to increase its probability of being selected as the leader. To address the above issues, the key idea behind PoW is that a participant will be randomly selected as the leader of each round with a probability in proportion to its computing power.  

In particular, blockchain implements PoW using computational hash puzzles. To create a new block, the nonce placed into the header of the block must be a solution to the hash puzzle expressed by the following inequality \cite{consensus2019}: 
\begin{equation}\label{poweuq}
bh = {\cal H}\left( {x\left\| {{{nonce}}} \right.} \right) \le T\left( d \right)
\end{equation} 
where $x$ denotes the binary string assembled using the candidate block data including the Merkle root of all transactions, the hash of the previous block, etc., $nonce$  denotes the solution string of the nonce, ${\cal H} \left(  \cdot  \right)$  is the cryptographic hash function, $bh$ is the hash of the candidate block and it is a bit stream of length $L$, $T\left( d \right) = {2^{L - d}}$ is a target value, $d$ is the current difficulty level of the hash puzzle (i.e., the number of leading zeros in the hash of a valid candidate block). Using the blockchain terminology, the process of computing hashes to find a nonce is called mining, and the participants involved are called miners.

\begin{figure}[!t]
	\centering
	\includegraphics[width=3.3in]{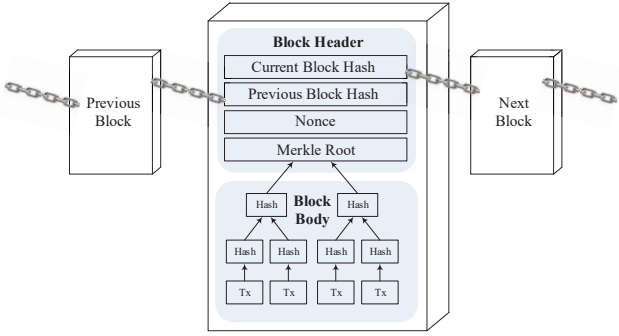}
	\caption{Illustration for the data structure of Bitcoin blockchain.} \label{blc}
\end{figure}

With a difficult level $d$ and the corresponding target $T\left( d \right) = {2^{L - d}}$ in (\ref{poweuq}), each single query to the PoW puzzle expressed in (\ref{poweuq}) is an i.i.d. Bernoulli test whose success probability is given by 
\begin{equation}
\Pr \left( {y:H\left( {x\left\| {\rm{y}} \right.} \right) \le T\left( d \right)} \right) = \frac{{T\left( d \right)}}{{{2^L}}} = {2^{ - d}}
\end{equation}
When $d$ is very large, the above success probability of a single query is very tiny. Moreover, it is known that with a  secure hash algorithm (e.g., the SHA-256 hash used for Bitcoin), the only way to solve (\ref{poweuq}) is to query a large number of nonces one by one to check if (\ref{poweuq}) is fulfilled until one lucky nonce is found (i.e., to exhaustively search for the nonce). Therefore, the probability of finding such nonce is proportional to the computing power of the participant---the faster the hash function in (\ref{poweuq}) can be computed in each trial, the more number of nonces can be tried per unit time.

Miners need to compute hash queries as fast as possible to win the race of mining, which is a very computationally intensive task. Let ${w_n}$ denote the number of hash queries that miner $n$ can compute per unit time, i.e., ${w_n}$ is the hash rate of miner $n$. Then, the number of success queries that miner $n$ can make converges to a Poisson process with rate ${\lambda _n} \buildrel \Delta \over = {w_n}/{2^d}$ \cite{consensus2019}. Moreover, the computation time between two successful queries made by miner $n$  (represented by a random variable ${X_n}$) fulfills the exponential distribution with rate  ${\lambda _n}$
   \cite{consensus2019}. Thus, the probability that at least one successful query made by miner $n$ within the duration of length $t$ is given by 
\begin{equation}
\Pr \left( {{X_n} < t} \right) = 1 - {e^{ - \frac{{{w_n}}}{{{2^d}}}t}}
\end{equation}
which is proportional to the hash rate ${w_n}$ and the mining-duration length $t$. It is evident from (3) that more computation power (faster hash rate) and more computation time (longer mining duration) lead to larger probability of successful mining.

Consider there are totally $N$  miners in the network and each performs mining to solve the PoW puzzle independently. Since the combination of the $N$ independent Poisson processes is still a Poisson process with a rate obtained by summing up the rates of the $N$ independent Poisson processes \cite{fisz2018probability}, the number of success queries per unit time made by the whole network is a Poisson processes with rate 
\begin{equation}
\lambda  = \sum\nolimits_{n = 1}^N {{\lambda _n} = \frac{1}{{{2^d}}}\sum\nolimits_{n = 1}^N {{w_n}} } 
\end{equation}
which is also the expectation of the successful queries made by the whole network per unit time. Therefore, the average number of blocks mined during the given block interval $T$ is $\lambda T = (T/{2^d})\sum\nolimits_{n = 1}^N {{w_n}} $. The difficulty control of blockchain aims at fixing the average number of the mined blocks per block interval $T$ to one by adjusting the difficulty level $d$ to adapt to the fluctuations in the total computation power of the network \cite{kraft2016difficulty}.

\subsection{Honest Mining Strategy}
When a miner tries to append a new block to the latest legal block by placing the hash of the latest block in the header of the new block, we say that the miner mines on the latest block. Bitcoin blockchain is maintained by miners in the following manner.

To encourage all miners to mine on (maintain) the current blockchain, each legal block distributes a reward to the miner as incentives. The reward of each block consists of two parts. The first part of the reward is a certain amount of new coins. When a miner mines a new block, the miner is allowed to place a coin-mint transaction in its mined block that credits this miner with some new coins as a part of the reward. The other part of the reward is the transaction fees contained in the transactions packaged in the block. If the block is verified and accepted by the blockchain network (i.e., it becomes a legal block), the reward is effective and thus can be spent on the blockchain. When a miner has found an eligible nonce, it publishes his block to the whole blockchain network. Other miners then verify the nonce and verify the transactions contained in that block. If the verification of the block is passed, other miners will mine on the block; otherwise, other miners discard the block and will continue to mine on the previous legal block. 

If two miners publish two different legal blocks that refer to the same preceding block at the same time, the blockchain is then forked into two branches. This is called forking of the blockchain. Forking is an undesirable feature of blockchain, since it threatens the security of blockchain \cite{bagaria2018deconstructing}. To resolve forking, PoW prescribes that only the rewards of blocks on the longest branch (called the main chain) are effective. Then, miners are incentivized to mine on the longest branch, i.e., miners always add new blocks after the last block on the longest main chain that is observed from their local perspectives. If the forked branches are of equal length, miners may mine subsequent blocks on either branch randomly. This is referred to as the rule of longest chain extension. 

The mining strategy of adhering to the rule of longest chain extension and publishing a block immediately after the block is mined is referred to as the honest mining strategy \cite{consensus2019}. The miners that comply with honest mining are called honest miners. It was widely believed that the most profitable mining strategy for miners is the honest mining strategy; and that when all miners adopt the honest mining strategy, each miner is rewarded proportionally to the ratio of its computing power to the total computing power all miners \cite{consensus2019}. As a result, any rational miner will not deviate from honest mining. This belief was later shown to be ill-founded for Bitcoin blockchain and that other mining strategies with higher profits are possible, such as selfish mining \cite{eyal2018majority}, withholding mining \cite{bag2016bitcoin}, etc.

To restrain forks on blockchain that will threaten the security, the generation rate of key blocks cannot be too small and the block size cannot too large \cite{decker2013information}. Therefore, the drawback of Bitcoin blockchain---its low TPS throughput---cannot be solved by just shortening the inter-block interval and inserting more transactions into each block.

\section{Bitcoin-NG Blockchain Protocol}

Compared with Bitcoin, Bitcoin-NG is a scalable blockchain protocol that allows for greater TPS throughputs without inducing extra communication latency. To separate the functionalities of selecting leaders (using PoW) and recording transactions, the Bitcoin-NG blockchain protocol introduces two types of blocks: key blocks and micro blocks. Bitcoin’s block and Bitcoin-NG’s key block have the same effectiveness expect that the latter contains no transactions. In Bitcoin-NG, the first miner that correctly solves the current hash puzzle creates a new key block and becomes the leader for the current round. After placing the new key block on the previous block, the leader is in charge of packing transactions into micro blocks. Intuitively, the Bitcoin-NG protocol divides a Bitcoin block into the key block and the micro block to achieve TPS throughput improvement.

\subsection{Key Blocks and Micro Blocks}
Fig. 2 illustrates the data structure of Bitcoin-NG blockchain. In each round (e.g., the $i$-th round), once a miner who finds a correct nonce to solve the PoW problem, this miner becomes the new leader and immediately creates a new key block ${K_i}$. Unlike the block of Bitcoin, this key block contains no transactions. It still contains the hash of the preceding block, the hash of this block, the nonce, a coin-base transaction to pay out the reward; moreover, it contains an extra public key. This public key must match the private key contained in the subsequent micro blocks.

After the key block, the miner generates many consecutive micro blocks, $\left\{ {{M_{i,1}},{M_{i,2}}, \cdots ,{M_{i,j}}, \cdots } \right\}$, that are used to pack transactions. Unlike the key block, the generation of these micro blocks does not need PoW. Thus, the leader can generate consecutive micro blocks quickly without extra computational overhead until the next key block is published. The header of each micro block encapsulates the hash of the preceding block, the Unix time, the hash of ledger entries and a signature of the header. The signature is signed with the private key that matches the public key contained in the key block ${K_i}$. By packing many transactions into each micro block and publishing micro blocks in a relatively high rate, Bitcoin-NG is allowed to achieve very high TPS throughputs \cite{eyal2016bitcoin}. In \cite{bagaria2018deconstructing}, it is also theoretically analyzed that Bitcoin-NG-like protocols, which decouple the functionalities of leader selection and transaction recording into different types of blocks, can achieve the optimal transaction processing capacity of the network.

\begin{figure}[!t]
	\centering
	\includegraphics[width=3.2in]{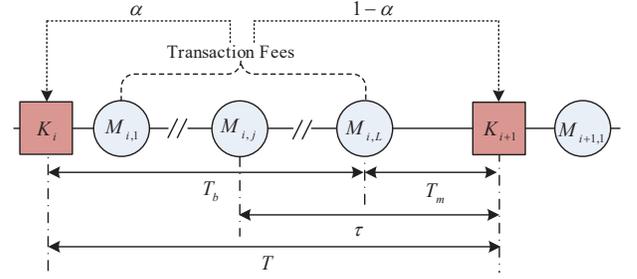}
	\caption{Illustration for the data structure of Bitcoin-NG blockchain. A key block (the square) is followed with a set of micro blocks (the circle). Transaction fees are divided into two parts: $\alpha$  for the current leader and  $1-\alpha$  for the next leader.} \label{mining_time}
\end{figure}

\subsection{Incentives }
Bitcoin-NG employs its specially designed incentive mechanism to motivate rational miners to follow the three honest actions: i) extending the heaviest chain; ii) extending the longest chain; iii) including transactions into the micro blocks. 

\textbf{Heaviest chain extension:} Assuming a majority of miners in the network are honest, Bitcoin-NG is designed to incentivize miners to always extend the heaviest chain that contains the largest amount of proof-of-work. In Bitcoin, the heaviest chain is the longest chain, since each block (that contains proof-of-work) is given a weight. In Bitcoin-NG, only key blocks are given weights and micro blocks are given no weight (since micro blocks contain no proof-of-work). Without assigning weights to micro blocks, Bitcoin-NG does not increase the system’s vulnerability to a kind of selfish mining \cite{eyal2018majority} where the leader of the current round will hide some micro blocks and mine on a hidden micro block privately.

\textbf{Longest chain extension:} In Bitcoin-NG, the transaction fees contained in the micro blocks are split to the two leaders: a fraction $\alpha$ of the transaction fees is released to the leader of the current round and a fraction $1-\alpha$ of the transaction fees is rewarded to the leader of the next round, as shown in Fig. 2. This is different from the incentive mechanism of Bitcoin that rewards the current leader with all transaction fees contained in the current block. This incentive mechanism of Bitcoin-NG can encourage miners to extend the longest chain. 

Suppose that all transaction fees contained in these micro blocks will be rewarded to the current leader. To earn more revenue, a malicious miner may deliberately discard the latest published micro block and mine on an earlier published micro block. If this miner succeeds in doing that, he will pack the transactions of the discarded micro block into his own micro blocks to get all transaction fees. Splitting the transaction fees into two parts can incentivize miners to mine on the longest chain that contains the already published micro blocks, since to become the next leader can still earn transaction fees from these micro blocks. 

However, even with this incentive mechanism, it is still possible that miners can earn more revenue by deliberately discarding the latest published micro block and mining on an earlier published micro block. The more possible revenue can be achieved using the following mining strategy. If one miner succeeds in mining the one key block, he will then pack the transactions (that contained in the previous micro block discarded by her/him) into his own micro block and continue to mine on the next key block. Hence, the value of  $\alpha$  should be designed such that the revenue of the leader taking this mining strategy must be smaller than his/her revenue of expanding the longest chain \cite{eyal2016bitcoin}.

\textbf{Transaction inclusion:} Moreover, the value of parameter $\alpha$ should be chosen to motivate the current leader to spontaneously pack transactions into its published micro block. Since the transaction fees are split to the current leader and the next leader, the current leader has a general incentive to pack transactions into a hidden micro block and mine on the hidden micro block to potentially obtain 100\% of the transaction fees. The value of $\alpha $ should to be chosen such that the leader's revenue of withholding the micro block must be smaller than his revenue of abiding by the protocol \cite{eyal2016bitcoin}.

Considering the above possible malicious mining behaviors, the value of $\alpha $ is suggested to $\alpha=0.4$ by the incentive mechanism in the original design of Bitcoin-NG \cite{eyal2016bitcoin}. However, the analysis on the incentive mechanism of Bitcoin-NG is flawed. Ref. \cite{yin2018revisiting} points out that there is a negligence and an over-simplification on the original analysis of Bitcoin-NG incentive mechanism, and it corrects the optimal value of $\alpha $ as $\alpha  = {3 \mathord{\left/
		{\vphantom {3 {11}}} \right.
		\kern-\nulldelimiterspace} {11}}$. Ref. \cite{niu2020incentive}  investigates the incentive mechanism of Bitcoin-NG by considering the selfish mining of key blocks and micro blocks jointly. In this work, we reveal that besides the above discussed malicious mining behaviors that are solved by the incentive mechanism of Bitcoin-NG, there still exists a possible malicious mining behavior in the Bitcoin-NG network.

\section{Game-Theoretical Analysis of Bitcoin-NG Mining}
In this section, we study a new identified malicious mining---advanced mining attack---for Bitcoin-NG blockchain. 

\subsection{One-Attacker Mining Optimization}
First, we consider the scenario where an attacker can change its mining-duration length to perform advanced mining attack to other honest miners who follow the rule of mining in a fixed default duration. We formulate the advanced mining attack problem as an optimization problem for the attacker to find the optimal length of the mining duration that maximizes the attacker's revenue.

It is theoretically shown in \cite{bagaria2018deconstructing} that blockchain protocols can achieve optimal transaction throughputs by decoupling each block into two kinds of blocks where one is used for selecting the leader and the other is used for recording transactions. In \cite{bagaria2018deconstructing}, such decouple is achieved by setting different mining targets: the mining target of blocks for recording transactions is smaller than that of blocks for selecting leaders. Bitcoin-NG is a special implementation of such blockchain protocol, i.e., Bitcoin-NG assigns mining target $T(d)=2^{L-d}$ to key blocks and no mining target to micro blocks. The mining target $T(d)=2^{L-d}$ is determined by the difficult control that aims to maintain the average interval between two consecutive key blocks at a constant value.

After the key block of the current round is published, the leader broadcasts consecutive micro blocks to the network within a duration of length ${T_b}$. Following the latest micro block, miners try to compute the next key block. We denote ${T_m}$ as the length of the duration between the last micro block of this round and the key block of the next round, and assume that all the honest miners adopt this duration as the default mining duration. Therefore, the length of the interval between two adjacent key blocks is given by ${T={T_b}+{T_m}}$.  Since ${T_m}$ is the length of the default mining duration, the difficulty control of Bitcoin-NG is made with respect to ${T_m}$, i.e., by adjusting the difficult level  $d$  to maintain $({T_m}/{2^d})\sum\nolimits_{n = 1}^N {{w_n}} = 1$. With this design of difficult control for Bitcoin-NG, we can fix the average interval between two adjacent key blocks to a constant.\footnote{In the original Bitcoin-NG protocol that follows the security designs of Bitcoin, the difficulty control is made with respect to $T$; and all miners mine immediately after the key block; then after each micro block is published by the current leader, all miners change to mine on the new micro block (like mining a new block in Bitcoin). However, this design of Bitcoin-NG cannot fix the average interval between two adjacent key blocks to a constant. This is because that when miners change to mine on a new micro block, the mining process of the next key block is restarted due to that the mining poison process is memoryless; then, the average time between this micro block and the next successful mined key block is still $T$.}

Although the default mining duration is preset, miners still can freely decide when to begin their mining due to the decentralized nature of the system. On one hand, since $1-\alpha $ of the transaction fees contained in the micro blocks of this round are distributed to the next leader, miners generally turn to mine on the latest published micro block to earn transaction fees as many as possible. On the other hand, if miners choose to mine on an early micro block, they will lose a part of the transaction fees but have more time to make more hash queries for computing the nonce. The greater number of nonces being tried admits a higher probability of finding the correct nonce. If an early miner succeeds to find the next key block earlier than other miners, the lost transaction fee can be compensated by the reward of the new coins mint in the next key block. Therefore, the possible revenue motivates miners to mine on an early micro block. We refer to such mining strategy as advanced mining attack.

We now consider that there is one attacker making advanced mining attack in the Bitcoin-NG network. During the interval between the current key block and the next key block, this attacker uses the last duration of length $\tau $ to compute the next key block, where $\tau  > {T_m}$. Without loss of generality, we group all other miners in the network as a single honest miner, who uses the last duration of length ${T_m}$ to compute the next key block according to the default mining rule. In Fig. 2, we illustrate the relationship among the variables of $T$, ${T_b}$, ${T_m}$ and  $\tau $ defined above. Let ${w_A}$ denote the hash rate of the attacker, and ${w_B}$ denote the hash rate of the honest miner. If the perfect difficulty control with respect to ${T_m}$ is achieved, we have ${T_m}({w_A} + {w_B})/{2^d} = 1$.

Actually, the attacker and the honest miner devote their computation powers to perform a mining race: the one who computes a valid nonce earlier than the other is the winner. The winner will be the leader of the next round and can earn the corresponding rewards. Therefore, devoting more computation resources (i.e., longer mining duration) can increase the winning chance. We denote the probability that the attacker wins the mining race by a function of the length of its mining duration $\tau$, $P\left( \tau  \right)$. Note that $P\left( \tau  \right)$ is monotonically increasing with respect to $\tau$, indicating that the attacker achieves a higher wining probability when a longer mining duration is devoted.

We denote the lengths of the mining duration needed for successful mining of the attacker and the honest miner as random variables ${X_A}$ and ${X_B}$, respectively. The random variables ${X_A}$ and ${X_B}$ are independent and both fulfill the exponential distribution with rates ${\lambda _A} \buildrel \Delta \over = w_A/2^d$ and ${\lambda _B} \buildrel \Delta \over = w_B/2^d$ \cite{kraft2016difficulty}. Hence, the winning probability $P\left( \tau  \right)$  of the attacker is given by 
\begin{equation}
\begin{gathered}
P\left( \tau  \right) = \Pr \left( {{X_A} - \tau  < {X_B} - {T_m}} \right) \\ 
\;\;\;\;\;\;\;\;\;= \Pr \left( {{X_A} - {X_B} < \tau  - {T_m}} \right) \\ 
\end{gathered} 
\end{equation}
where ${{X_A} - \tau}$ and ${{X_B} - {T_m}}$  are the time instances when the attacker and the honest miner successfully find the next key block. Since the difference between two independent exponential distributed random variables, e.g., ${{X_A} - {X_B}}$, is a Laplace distributed random variable, the probability expressed in (5) can be calculated as \cite{yin2018revisiting}:
\begin{equation}
P\left( \tau  \right) = 1 - \frac{{{{\rm{\lambda }}_B}}}{{{{\rm{\lambda }}_{\rm{A}}} + {{\rm{\lambda }}_{\rm{B}}}}}{e^{ - \left( {\tau  - {T_m}} \right){{\rm{\lambda }}_{\rm{A}}}}}
\end{equation}
If we consider that there is no advanced mining attack, i.e., ${\tau= {T_m}}$, the probability in (6) is reduced to $P\left( {\tau  = {T_m}} \right) = {{{{\rm{\lambda }}_A}} \mathord{\left/
		{\vphantom {{{{\rm{\lambda }}_A}} {\left( {{{\rm{\lambda }}_{\rm{A}}} + {{\rm{\lambda }}_{\rm{B}}}} \right)}}} \right.
		\kern-\nulldelimiterspace} {\left( {{{\rm{\lambda }}_{\rm{A}}} + {{\rm{\lambda }}_{\rm{B}}}} \right)}}$, which is the successful probability of honest mining that equals the ratio of the miner's computation power over the totally computation power of the network \cite{consensus2019}.

We assume that each key block contains a reward $R$ (i.e., the value of the mint coins), each micro block contains a reward $r$ (i.e., the value of the transaction fees), and the current leader can generate $L$ micro blocks after its key block $K_{i}$ within the duration of length ${T_b}$. If the advanced mining attacker can mine the next key block $K_{i+1}$, it can earn the transaction fees from the first ${{L\left( {T - \tau } \right)} \mathord{\left/
		{\vphantom {{L\left( {T - \tau } \right)} {{T_b}}}} \right.
		\kern-\nulldelimiterspace} {{T_b}}}$  micro blocks, where ${{L\left( {T - \tau } \right)} \mathord{\left/
		{\vphantom {{L\left( {T - \tau } \right)} {{T_b}}}} \right.
		\kern-\nulldelimiterspace} {{T_b}}<L}$  since $T - \tau  < {T_b}$. Therefore, after the key block $K_{i+1}$ is published by the attacker, the total reward obtained by the attacker in the last block interval includes both the rewards from the new key block $K_{i+1}$ and the rewards from the micro blocks $\left\{ {{M_{i,1}},{M_{i,2}}, \cdots ,{M_{i,{{L\left( {T - \tau } \right)} \mathord{\left/
					{\vphantom {{L\left( {T - \tau } \right)} {{T_b}}}} \right.
					\kern-\nulldelimiterspace} {{T_b}}}}}} \right\}$ that are included onto the blockchain. When computing the total reward for the attacker, we consider two different situations: 1) the previous key block $K_{i}$ is mined by the honest miner (i.e., ${K_{i}} \in {B_{honest}}$); 2) the previous key block $K_{i}$ is mined by the attacker (i.e., ${K_{i}} \in {B_{attacker}}$), where ${B_{attacker}}$ (${B_{honest}}$) is the set of the key blocks mined by the attacker (honest miners).

We first consider that the previous key block $K_{i}$ is mined by the honest miner (${K_{i}} \in {B_{honest}}$). If the attacker can succeed to find the next key block $K_{i+1}$, the total reward of the attacker includes both the rewards from the new key block and the $\left( {1 - \alpha } \right) $-fraction of the rewards from the micro blocks, i.e., the total reward is given by:
\begin{equation}
Q\left( \tau  \right) = \left( {1 - \alpha } \right)rL\frac{{T - \tau }}
{{{T_b}}} + R
\end{equation}
which is also a function of its mining-duration length. The expected reward of the attacker (denoted by $\pi \left( \tau  \right)$) is the total revenue when its mining is successful $Q\left( \tau  \right)$ multiplied by the probability of successful mining $P\left( \tau  \right)$:
\begin{equation}
\begin{gathered}
\pi \left( \tau  \right) = Q\left( \tau  \right)P\left( \tau  \right) \hfill \\
= \left( {\left( {1 - \alpha } \right)rL\frac{{T - \tau }}
	{{{T_b}}} + R} \right)\left( {1 - \frac{{{\lambda _B}}}
	{{{\lambda _A} + {\lambda _B}}}{e^{ - \left( {\tau  - {T_m}} \right){\lambda _A}}}} \right) \hfill \\ 
\end{gathered} 
\end{equation}
The reward function $\pi \left( \tau  \right)$  is concave and continuous in $\tau$ , and we can obtain the maximum value of $\pi \left( \tau  \right)$  when
\begin{equation}\label{cond1}
\frac{{d\pi (\tau )}}{{d\tau }} = 0
\end{equation}
Solving (\ref{cond1}), we find that the reward function $\pi \left( \tau  \right)$  achieves its maximum value at
\begin{equation}\label{op1}
\begin{gathered}
{\tau ^*} =  \hfill \\
\frac{{\frac{R}
		{{\left( {1 - \alpha } \right)rL}}{\lambda _A}{T_b} + {\lambda _A}T + 1 - W\left( {\frac{{{1}}}
			{{1 - {\lambda _A}{T_m}}}{e^{\left( {\frac{R}
						{{\left( {1 - \alpha } \right)rL}} + 1} \right){\lambda _A}{T_b} + 1}}} \right)}}
{{{\lambda _A}}} \hfill \\ 
\end{gathered} 
\end{equation}
where $W\left(  \cdot  \right)$  denotes the Lambert W Function \cite{corless1996lambertw}.

We then consider that the previous key block $K_{i}$ is also mined by the attacker (${K_{i}} \in {B_{attacker}}$). Now, if the attacker can succeed to find the next key block $K_{i+1}$, the total reward of the miner includes both the rewards from the new key block and all the rewards from the micro blocks:
\begin{equation}
Q\left( \tau  \right) = rL\frac{{T - \tau }}
{{{T_b}}} + R
\end{equation}
Following the same way, we can compute the optimal mining length that achieves the maximum mining reward as: 
\begin{equation}\label{op2}
\begin{gathered}
{\tau ^*} = \frac{{\frac{R}
		{{rL}}{\lambda _A}{T_b} + {\lambda _A}T + 1 - W\left( {\frac{{{1}}}
			{{1 - {\lambda _A}{T_m}}}{e^{(\frac{R}
					{{rL}} + 1){\lambda _A}{T_b} + 1}}} \right)}}
{{{\lambda _A}}} \hfill \\
\end{gathered} 
\end{equation}
According to whether the previous key block is mined by the attacker, the attacker can choose its optimal mining length $\tau ^*$ according to (\ref{op1}) or (\ref{op2}). In Section V, we numerically compute the corresponding maximum expected reward $\pi \left( {{\tau ^*}} \right)$ that is showed to be larger than the expected reward earned by the honest mining $\pi \left( {{T_m}} \right)$. The result indicates that when all other miners adopt honest mining, the attacker's advanced mining over the last duration ${\tau ^ * } > {T_m}$ of the block interval is the optimal strategy to earn the highest expected reward.

\subsection{Two-Player Mining Game}
We next proceed to analyze the scenario where all miners in the network are divided into two mining pools that both try to make advanced mining attack (i.e., changing their mining-duration lengths) and thus compete with each other. We will see that this setup leads to a game-theoretical model and we derive the Nash equilibrium mining strategies for the two mining pools. We denote the two mining pools as pool $A$ and pool $B$  with hash rates ${w_A}$ and ${w_B}$, respectively. Pool $A$ ($B$) attempts to carry out advanced mining using the mining duration of length $\tau_A$ ($\tau_B$). The mining behaviors of the two pools in this scenario can be analyzed through a two-player game.

We formulate the two-pool advanced mining problem as a two-player game as follows. The two players, pool  $A$ and pool $B$, strategically choose their mining-duration lengths to compete for the reward of successful mining. The two mining pools are rational and their interaction can be modeled as a non-cooperative game \cite{abdalzaher2016game}. Each pool has a set of pure strategies in $S = [{T_m},T]$. Let ${\tau _n} \in S$  be the mining strategy of pool $n$, where $n \in \left\{ {A,B} \right\}$. A two-tuple of strategies of the two mining pools is ${\bf{\tau }} = \left( {{\tau _A},{\tau _B}} \right)$  and a two-tuple of corresponding payoffs is ${\bf{\pi }} = \left( {{\pi _A}\left( {\bf{\tau }} \right),{\pi _B}\left( {\bf{\tau }} \right)} \right)$, where  ${\pi _n}\left( {\bf{\tau }} \right)$ is the utility of player $n$ given the chosen strategies of the two mining pools. Each mining pool chooses its best strategy $\tau _n^*$ to maximize its utility. A set of strategies ${{\bf{\tau }}^*} = \left( {\tau _A^*,\tau _B^*} \right)$ is the Nash equilibrium if no miner can gain higher utility by unilaterally changing its own strategy when the strategies of the other miners remain unchanged, i.e., 
\begin{equation}\label{cond2}
\forall {\bf{\tau }} = \left( {{\tau _A},{\tau _B}} \right) \in S \times S:\left\{\begin{aligned}
{\pi _A}\left( {\tau _A^*,\tau _B^*} \right) \ge {\pi _A}\left( {{\tau _A},\tau _B^*} \right)\\
{\pi _B}\left( {\tau _A^*,\tau _B^*} \right) \ge {\pi _B}\left( {\tau _A^*,{\tau _B}} \right)
\end{aligned}
\right.\end{equation}
The inequalities in (\ref{cond2}) defines the equilibrium state of the game. At the Nash equilibrium if it exists, the players have no incentive to deviate from their equilibrium strategies.

For the advanced mining problem, we adopt the expected mining rewards as the utilities in the game formulation. We compute the utilities as follows. After the two mining pools have successful mined the next key block $K_{i+1}$, their rewards are 
\begin{equation}\label{reward21}
\left\{ \begin{gathered}
{Q_A}\left( {{\tau _A}} \right) = \left( {1 - \alpha } \right)rL\frac{{T - {\tau _A}}}
{{{T_b}}} + R \hfill \\
{Q_B}\left( {{\tau _B}} \right) = rL\frac{{T - {\tau _B}}}
{{{T_b}}} + R \hfill \\ 
\end{gathered}  \right.
\end{equation}
if the previous block $K_{i}$ is mined by mining pool $B$ (${K_{i}} \in {B_B}$), or 
\begin{equation}\label{reward22}
\left\{ \begin{gathered}
{Q_A}\left( {{\tau _A}} \right) = rL\frac{{T - {\tau _A}}}
{{{T_b}}} + R \hfill \\
{Q_B}\left( {{\tau _B}} \right) = \left( {1 - \alpha } \right)rL\frac{{T - {\tau _B}}}
{{{T_b}}} + R \hfill \\ 
\end{gathered}  \right.
\end{equation}
if the previous block $K_{i}$ is mined by mining pool $A$ (${K_{i}} \in {B_A}$).

Only if the time instance for a pool finding a nonce is earlier than the time instance for its opponent does, that mining pool can earn the reward. Since the mining pools both can change the lengths of their mining duration, ${\tau_A}$ and ${\tau_B}$, the successful mining probability of pool $n$ is a function of mining-duration lengths of both pools, i.e., ${\bf{\tau }} = \left( {{\tau _A},{\tau _B}} \right)$. We write the successful mining probabilities of pool $A$ and pool $B$ as follows 
\begin{equation}\label{prob22}
\left\{\begin{aligned}
{P_A}\left( {\bf{\tau }} \right) &= \Pr \left( {{X_A} - {\tau _A} < {X_B} - {\tau _B}} \right)\\ &= \Pr \left( {{X_A} - {X_B} < {\tau _A} - {\tau _B}} \right)\\
{P_B}\left( {\bf{\tau }} \right) &= \Pr \left( {{X_B} - {\tau _B} < {X_A} - {\tau _A}} \right)\\ &= \Pr \left( {{X_B} - {X_A} < {\tau _B} - {\tau _A}} \right)
\end{aligned}
\right.\end{equation}
where ${X_A}$  and ${X_B}$  are the random variables representing the lengths of the mining duration needed for the successful mining of mining pools $A$ and $B$, respectively. We find that the computation of probabilities ${P_A}\left( {\bf{\tau }} \right)$ and ${P_B}\left( {\bf{\tau }} \right)$  depends on the sign of ${\tau _A} - {\tau _B}$, and thus we can analyze the  cases of ${\tau _A} - {\tau _B}<0$  and ${\tau _A} - {\tau _B}>0$, respectively. 

We derive the Nash equilibrium for the two-player mining game when ${\tau _A} - {\tau _B}<0$ in the following. The differences of two independent exponential distributed random variables, ${X_A}-{X_B}$, are Laplace distributed. When ${\tau _A} - {\tau _B}<0$, the probabilities ${P_A}\left( {\bf{\tau }} \right)$ and ${P_B}\left( {\bf{\tau }} \right)$ in (\ref{prob22}) can be computed as: 
\begin{equation}\label{prob222}
\left\{\begin{aligned}
{P_A}\left( {\bf{\tau }} \right) &= \frac{{{\lambda _A}}}{{{\lambda _A} + {\lambda _B}}}{e^{\left( {{\tau _A} - {\tau _B}} \right){\lambda _B}}}
\\
{P_B}\left( {\bf{\tau }} \right) &= 1 - \frac{{{\lambda _A}}}{{{\lambda _A} + {\lambda _B}}}{e^{ - \left( {{\tau _B} - {\tau _A}} \right){\lambda _B}}}
\end{aligned}
\right.\end{equation}
where ${\lambda _A} = w_A/2^d$  and ${\lambda _B} = w_B/2^d$. 

We assume that the previous block $K_{i}$ is mined by mining pool $B$ (${K_{i}} \in {B_B}$). Using (\ref{reward21}) and (\ref{prob222}), the utilities of the two players (i.e., their expected mining rewards) are given by 
\begin{equation}
\left\{\begin{aligned}
{\pi _A}\left( {\bf{\tau }} \right) &= {Q_A}\left( {{\tau_A }} \right){P_A}\left( {{\tau}} \right)\\
&= 
(\left( {1 - \alpha } \right) rL\frac{{T - {\tau _A}}}{{{T_b}}} + R)\frac{{{\lambda _A}}}{{{\lambda _A} + {\lambda _B}}}{e^{\left( {{\tau _A} - {\tau _B}} \right){\lambda _B}}}\\
{\pi _B}\left( {\bf{\tau }} \right) &= {Q_B}\left( {{\tau_B }} \right){P_B}\left( {{\tau}} \right)\\
&= 
( rL\frac{{T - {\tau _B}}}{{{T_b}}} + R)(1 - \frac{{{\lambda _A}}}{{{\lambda _A} + {\lambda _B}}}{e^{\left( {{\tau _A} - {\tau _B}} \right){\lambda _B}}})
\end{aligned}
\right.\end{equation}
The utilities ${\pi _A}\left( {\bf{\tau }} \right)$ and ${\pi _B}\left( {\bf{\tau }} \right)$  are concave and continuous in ${\tau _A}$ and ${\tau _B}$, and thus the Nash equilibrium must satisfy the condition 
\begin{equation}\label{cond2222}
\left\{ {\frac{{\partial {\pi _A}\left( {{\tau _A},{\tau _B}} \right)}}{{\partial {\tau _A}}} = 0,\frac{{\partial {\pi _B}\left( {{\tau _A},{\tau _B}} \right)}}{{\partial {\tau _B}}} = 0} \right\}
\end{equation}
Solving (\ref{cond2222}), we obtain the Nash equilibrium for the two-player mining game when ${\tau _A} - {\tau _B}<0$ and ${K_{i}} \in {B_B}$: 
\begin{equation} \label{nash21}
\left\{ \begin{array}{l}
\tau _A^* = T + \frac{R}{{\left( {1 - \alpha } \right)rL}}{T_b} - \frac{1}{{{\lambda _A}}} \\ 
\tau _B^* = \frac{{\frac{R}{{rL}}{\lambda _B}{T_b} + {\lambda _B}T + 1 - W\left( {\frac{1}{{1{\rm{ - }}{\lambda _B}{T_m}}}{e^{2 - \frac{R}{{rL}}{\lambda _B}{T_b}\frac{\alpha }{{1 - \alpha }}}}} \right)}}{{{\lambda _B}}} \\ 
\end{array} \right.
\end{equation}
where we have used the difficulty control result ${T_m}\left( {{\lambda _A} + {\lambda _B}} \right) = 1$. Similarly, when ${\tau _A} - {\tau _B}<0$ and the previous block $K_{i}$ is mined by mining pool $A$ (${K_{i}} \in {B_A}$), the Nash equilibrium for the two-player mining game is: 
\begin{equation}\label{nash22}
\left\{ \begin{array}{l}
\tau _A^* = T + \frac{R}{{rL}}{T_b} - \frac{1}{{{\lambda _B}}} \\ 
\tau _B^* = \frac{{\frac{R}{{(1 - \alpha )rL}}{\lambda _B}{T_b} + {\lambda _B}T + 1 - W\left( {\frac{1}{{1{\rm{ - }}{\lambda _B}{T_m}}}{e^{2{\rm{ + }}\frac{R}{{rL}}{\lambda _B}{T_b}\frac{\alpha }{{1 - \alpha }}}}} \right)}}{{{\lambda _B}}} \\ 
\end{array} \right.
\end{equation}

For the case of ${\tau _A} - {\tau _B}>0$, the derivation for the Nash equilibrium is similar to the case of  ${\tau _A} - {\tau _B}<0$. We omit it here to save space. The mining lengths achieved at the Nash equilibrium indicate that for all cases, the two mining pools will deviate from mining with the default mining-duration length.

\subsection{$N$-Player Mining Game}
We finally extend the two-player mining game to a $N$-player mining game. We consider that the miners in the network are grouped into $N$ mining pools and all mining pools execute advanced mining. Mining pool $n \in \left\{ {1,2, \cdots ,N} \right\}$ with hash rates ${w_n}$ uses the mining duration of length ${\tau_n}$  to carry out advanced mining. The mining behaviors of the $N$  pools can also be modeled as a non-cooperative game. 

In the $N$-player game, all $N$  mining pools strategically choose the lengths of their mining duration to maximize their revenue. Each pool $n$  has a set of pure strategies $S = [{T_m},T]$ . Let ${\tau _n} \in S$  be the strategy of mining pool $n$.  The vector of the strategies of $N$  mining pools is ${\bf{\tau }} = \left( {{\tau _1},{\tau _2}, \cdots ,{\tau _N}} \right)$ , and the vector of the corresponding utilities is ${\bf{\pi }} = \left( {{\pi _1}\left( {\bf{\tau }} \right),{\pi _2}\left( {\bf{\tau }} \right), \cdots ,{\pi _N}\left( {\bf{\tau }} \right)} \right)$, where ${\pi _n}\left( {\bf{\tau }} \right)$  is the utility of player $n$  given the chosen strategies of the $N$  mining pools. Each mining pool chooses its best strategy $\tau _n^*$  to maximize its utility. The vector of the strategies  ${{\bf{\tau }}^*} = \left( {\tau _1^*,\tau _2^*, \cdots ,\tau _N^*} \right)$ is the Nash equilibrium if no mining pool can gain higher utility by changing its own strategy when the strategies of the other miners remain unchanged, i.e., 
\begin{equation}
\forall n \in \left\{ {1,2, \cdots ,N} \right\},\forall {\tau _n} \in S:{\pi _n}\left( {\tau _n^*,{{{\bf{\bar \tau }}}_n}^{\rm{*}}} \right) \ge {\pi _n}\left( {{\tau _n},{{{\bf{\bar \tau }}}_n}^{\rm{*}}} \right)
\end{equation}
where ${\bf{\bar \tau }}_n^* = (\tau _1^*, \cdots ,\tau _{n - 1}^*,\tau _{n + 1}^* \cdots ,\tau _N^*)$  is the vector of the best strategies of the other $N-1$  mining pools except mining pool $n$.

Then we compute the utility of pool $n$ in this game as follows. When we compute it, we assume the previous key block  $K_{i}$ is mined by the mining pool $N$ (the results for other $N-1$ cases can be obtained similarly). If mining pool $n$ succeeds to find the next key block  $K_{i+1}$ via advanced mining with mining duration of length ${\tau_n}$, its reward is given by
\begin{equation}\label{rewardn}
{Q_n}\left( {{\tau _n}} \right) = \left\{ {\begin{array}{*{20}{c}}
	{\left( {1 - \alpha } \right)rL\frac{{T - {\tau _n}}}
		{{{T_b}}} + R,} & {n = 1,2, \cdots ,N - 1}  \\
	{rL\frac{{T - {\tau _n}}}
		{{{T_b}}} + R,} & {n = N}  \\
	
	\end{array} } \right.
\end{equation}
Since the successful mining probability of mining pool $n$ is impacted not only by its own strategy but also other competitors' strategies, it can be written as a function of ${\bf{\tau }} = \left( {{\tau _1},{\tau _2}, \cdots ,{\tau _N}} \right)$  and computed as 
\begin{equation}\label{probn}
\begin{gathered}
{P_n}\left( \tau  \right) \hfill \\
= \Pr \left( {{X_n} - {\tau _n} < {X_1} - {\tau _1}, \cdots ,} \right.{X_n} - {\tau _n} < {X_{n - 1}} - {\tau _{n - 1}}, \hfill \\
\left. {{X_n} - {\tau _n} < {X_{n + 1}} - {\tau _{n + 1}}, \cdots ,{X_n} - {\tau _n} < {X_N} - {\tau _N}} \right) \hfill \\
= \prod\limits_{m \in \left\{ {1, \cdots n - 1,n + 1, \cdots ,N} \right\}} {\Pr \left( {{X_n} - {X_m} < {\tau _n} - {\tau _m}} \right)}  \hfill \\ 
\end{gathered} 
\end{equation}
where $X_n$ is the random variable representing the length of the mining duration needed for the successful mining of pool $n$. To compute the probabilities ${P_n}\left( {\bf{\tau }} \right)$ for all $n \in \left\{ {1,2, \cdots ,N} \right\}$ according to (\ref{probn}), we need to consider $\left( \begin{array}{l}
N\\
2
\end{array} \right)$ cases for the signs of  $\left\{ {{\tau _n} - {\tau _m},n = 1, \cdots ,N,m = n + 1, \cdots ,N} \right\}$.  For example, considering the case of $N=3$ and $\left\{ {{\tau _1} - {\tau _2} < 0,{\tau _1} - {\tau _3} < 0,{\tau _2} - {\tau _3} < 0} \right\}$, we have
\begin{equation}
\left\{\begin{aligned}
{P_1}\left( {\bf{\tau }} \right) &= \Pr \left( {{X_1} - {X_2} < {\tau _1} - {\tau _2}} \right)\Pr \left( {{X_1} - {X_3} < {\tau _1} - {\tau _3}} \right)\\
&= \left( {\frac{{{\lambda _1}}}{{{\lambda _1} + {\lambda _2}}}{e^{\left( {{\tau _1} - {\tau _2}} \right){\lambda _2}}}} \right)\left( {\frac{{{\lambda _1}}}{{{\lambda _1} + {\lambda _3}}}{e^{\left( {{\tau _1} - {\tau _3}} \right){\lambda _3}}}} \right)\\
{P_2}\left( {\bf{\tau }} \right) &= \Pr \left( {{X_2} - {X_1} < {\tau _2} - {\tau _1}} \right)\Pr \left( {{X_2} - {X_3} < {\tau _2} - {\tau _3}} \right)\\
&= \left( {1 - \frac{{{\lambda _1}}}{{{\lambda _1} + {\lambda _2}}}{e^{  \left( {{\tau _1} - {\tau _2}} \right){\lambda _2}}}} \right)\left( {\frac{{{\lambda _2}}}{{{\lambda _2} + {\lambda _3}}}{e^{\left( {{\tau _2} - {\tau _3}} \right){\lambda _3}}}} \right)\\
{P_3}\left( {\bf{\tau }} \right) &= \Pr \left( {{X_3} - {X_1} < {\tau _3} - {\tau _1}} \right)\Pr \left( {{X_3} - {X_2} < {\tau _3} - {\tau _2}} \right)\\
&= \left( {1 - \frac{{{\lambda _1}}}{{{\lambda _1} + {\lambda _3}}}{e^{  \left( {{\tau _1} - {\tau _3}} \right){\lambda _3}}}} \right)\left( {1 - \frac{{{\lambda _2}}}{{{\lambda _2} + {\lambda _3}}}{e^{  \left( {{\tau _2} - {\tau _3}} \right){\lambda _3}}}} \right)
\end{aligned}
\right.
\end{equation}

Using (\ref{rewardn}) and (\ref{probn}), the utility of mining pool $n$ can be expressed as 
\begin{equation}
{\pi _n}\left( {\bf{\tau }} \right) = {Q_n}\left( {{\tau _n}} \right){P_n}\left( {\bf{\tau }} \right)
\end{equation}
for all $n$. Then, we can find the Nash equilibrium ${{\bf{\tau }}^*} = \left( {\tau _1^*,\tau _2^*, \cdots ,\tau _N^*} \right)$  by solving the equation system 
\begin{equation}
\left\{ {\frac{{\partial {\pi _1}\left( \tau  \right)}}{{\partial {\tau _1}}} = 0, \cdots ,\frac{{\partial {\pi _n}\left( \tau  \right)}}{{\partial {\tau _n}}} = 0, \cdots ,\frac{{\partial {\pi _N}\left( \tau  \right)}}{{\partial {\tau _N}}} = 0} \right\}
\end{equation}
We can find that the utility functions showed in (18) for the two-player game and in (26) for the $N$-player game are all strictly concave. Therefore, we can conclude that the Nash equilibrium achieved in the advanced mining game is unique \cite{game0}. However, we cannot derive the explicit form of the Nash equilibrium for the  $N$-player game. In the next section, we will solve the Nash equilibrium in different cases using  numerical computations.

\section{Numerical Computations and System Simulations}

In this section, we present the results of numerical computations and system simulations to investigate the advanced mining problem for the Bitcoin-NG protocol.

\subsection{Numerical Computations}

We first numerically compute the analytical results provided in section IV. In our numerical computations and simulations, we set $T = 10$, ${T_m} = 2$, $L = 10$, $\alpha  = 3/11$, $r = 1$ and $R{\text{ }} \in \left\{ {1,5,10} \right\}$. 

\begin{figure}[t]
	\centering
	\includegraphics[width=3.1in]{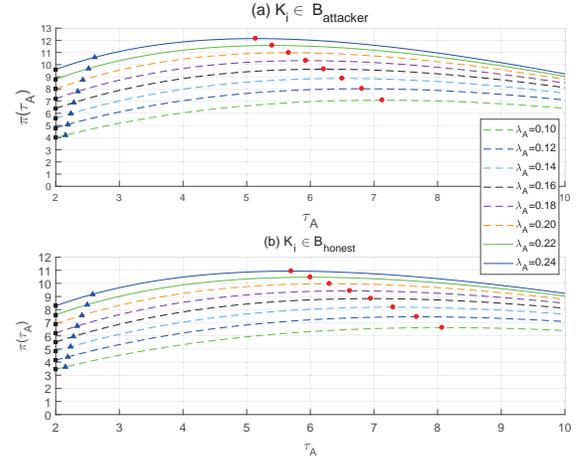}
	\caption{The expected reward of the attacker $\pi \left( \tau  \right)$ in the one-attacker mining optimization problem are given as the functions of the mining-duration length $\tau$.}
	\label{fig_sim_op}
\end{figure}

We first numerically analyze the one-attacker mining optimization problem. We compute the expected reward of the attacker $\pi \left( \tau  \right)$ that is a function of the length of the mining duration $\tau$, given the mining power ${\lambda _A}$ as the parameter. Note that since $T_m=2$, we have ${\lambda _A} + {\lambda _B} = {1 \mathord{\left/
		{\vphantom {1 {{T_m}}}} \right.
		\kern-\nulldelimiterspace} {{T_m}}} = 0.5$.  Fig. \ref{fig_sim_op} (a) presents the reward functions for the case of ${K_{i - 1}} \in {B_{honest}}$; and Fig. \ref{fig_sim_op} (b) presents the reward function for the case of ${K_{i}} \in {B_{attacker }}$. We set to $R=10$ when computing the expected reward. For a specific mining power, each red round represents the maximum expected reward achieved by the corresponding optimal mining-duration length  ${\tau ^ * }$ expressed in (\ref{op1}) for ${K_{i }} \in {B_{honest }}$ and in (\ref{op2}) for ${K_{i}} \in {B_{attacker }}$; each black square represents the expected reward achieved by the default mining-duration length ${T_m}$. From Fig. \ref{fig_sim_op}, we can see that higher mining powers decrease the required mining-duration lengths to achieve the maximum expected rewards. Intuitively, higher mining powers increase the probabilities of finding a key block, thus the attacker can mine slightly later and wait for more micro blocks to earn more transaction fees. Higher mining power also increase the maximum expected reward of the attacker. It is also observed that the optimal mining-duration length for ${K_{i }} \in {B_{honest }}$ is larger than that for ${K_{i }} \in {B_{attacker }}$. Since if the previous key block is mined by the attacker (${K_{i}} \in {B_{attacker }}$), the attacker is prone to include more micro blocks onto blockchain to earn the transaction fees and thus it reduces its mining-duration length. Moreover, we can observe that, whatever is the mining power, the maximum expected reward achieved by advanced mining is higher than that achieved by honest mining. 
	
For this one-attacker scenario, we also compare the advanced mining attack with the block withhold (BWH) attack with power splitting \cite{luu2015power}. BHW attack with power splitting for Bitcoin aims at gaining higher mining rewards by splitting the mining power of the attacker pool into multiple subparts and using some subparts to cancel the mining powers of other mining pools and some subparts to mine for itself. For comparison purpose, we apply BHW attack with power splitting to Bitcoin-NG. The BWH attacker adopts the optimal power splitting strategy derived in  \cite{luu2015power} for attacking and fixes its mining length to the default length ${T_m}$. We can compute the expected rewards achieved by BHW attack according to the result given in \cite{luu2015power}. In Fig. 3, each blue triangle represents the expected reward achieved by BWH attack for a specific mining power. We can observe that the achieved expected rewards of BHW attack are indeed higher than that of honest mining; the maximum expected rewards achieved by advanced mining are higher than that achieved by BWH attack. This indicates that in Bitcoin-NG blockchain, advanced mining attack is more effective than BWH attack.

\begin{figure*}[t]   
	\begin{minipage}[t]{0.33\linewidth} 
		\centering   
		\includegraphics[width=2.55in]{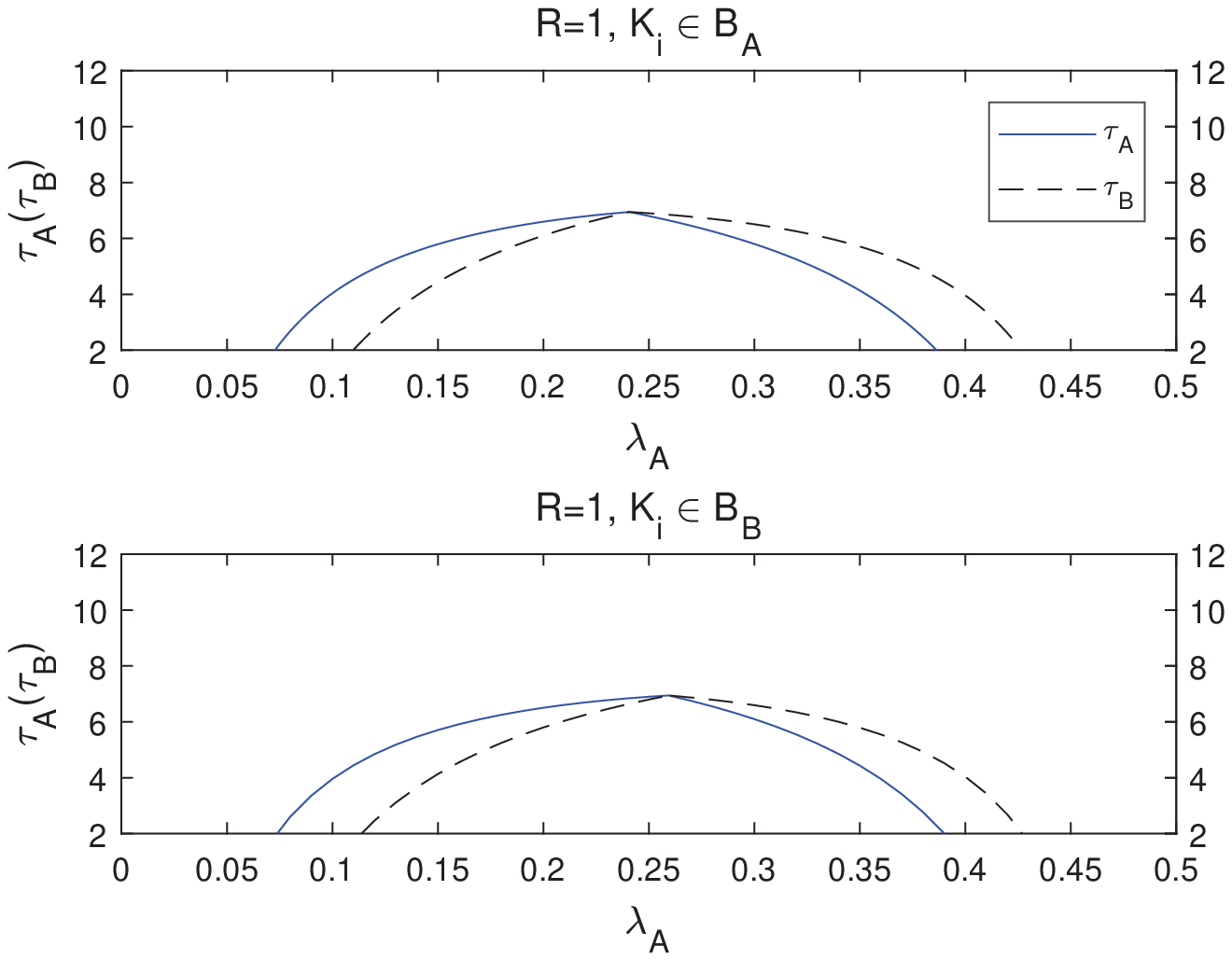}   
	\end{minipage}%   
	\begin{minipage}[t]{0.33\linewidth}   
		\centering   
		\includegraphics[width=2.55in]{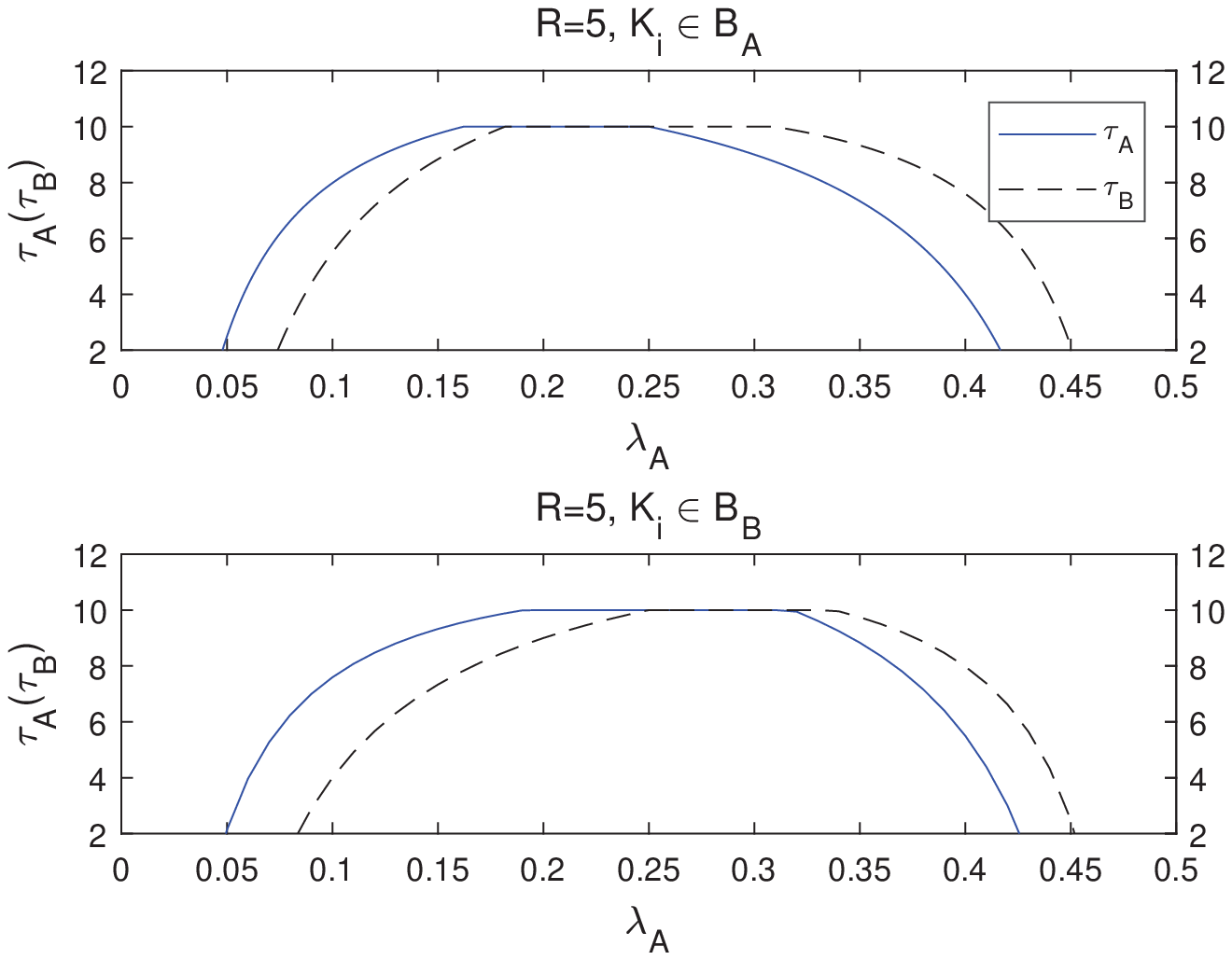}   
	\end{minipage}   
	\begin{minipage}[t]{0.33\linewidth}   
		\centering   
		\includegraphics[width=2.55in]{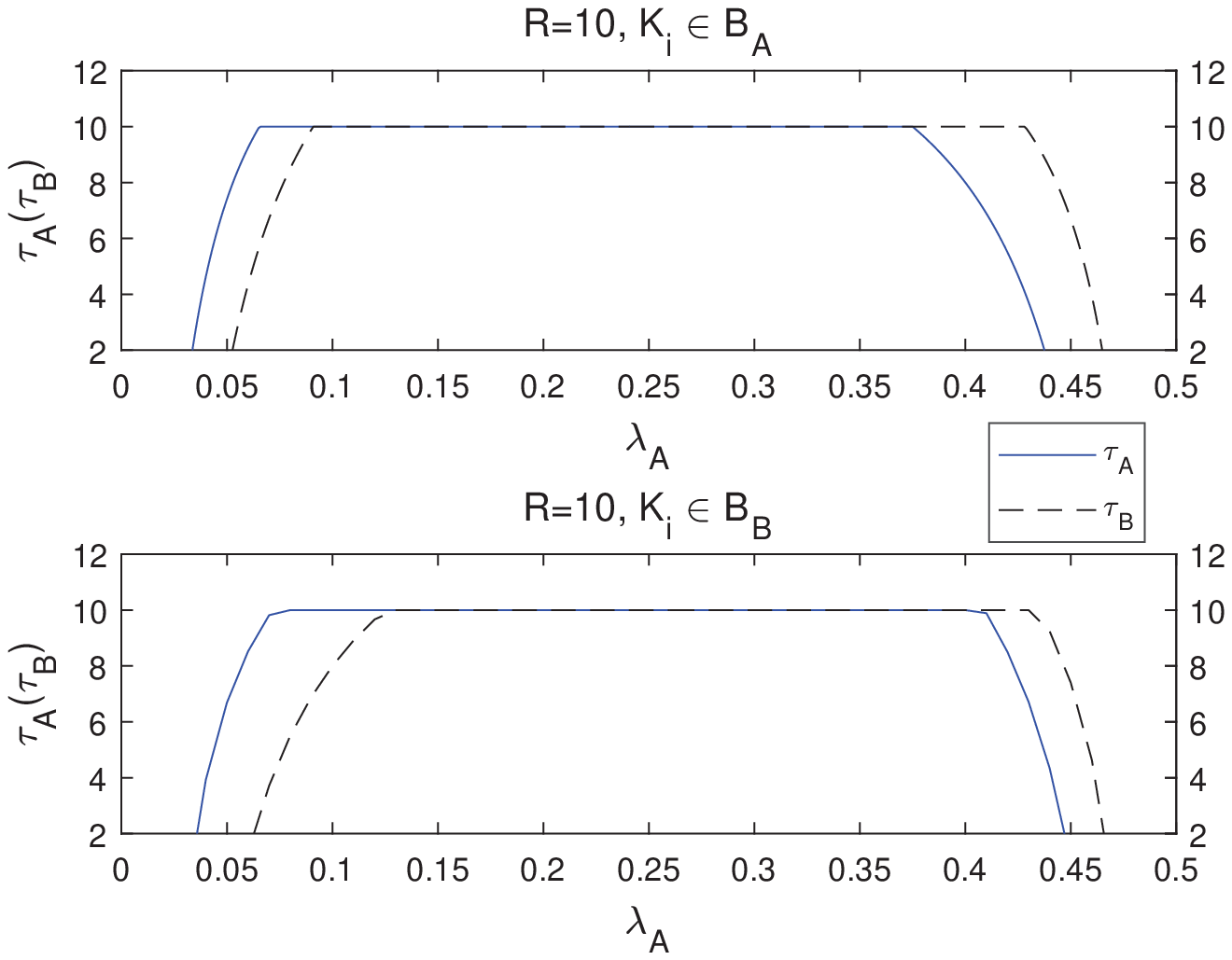}   
	\end{minipage}  
	\caption{The optimal mining-duration lengths of the two mining pools in the mining game model are given as functions of the mining power ${\lambda _A}$ ($R=\{1,5,10\}$).}  
	\label{fig_nu_two}   
\end{figure*} 

We then investigate the two-player mining game. We compute the optimal mining strategies (i.e., the optimal mining-duration lengths derived in Section IV.B). Fig. \ref{fig_nu_two} depicts the optimal mining-duration lengths of the two mining pools that are given as the functions of pool $A$'s mining power for $R=1,5,10$, respectively. We can see that when ${\lambda _A} < {\lambda _B}$, the optimal mining length of pool $A$ is larger than that of pool $B$. This fulfills the intuition that to achieve an equilibrium, the mining pool with less mining power needs to mine earlier to enlarge its successful mining probability. The situation is the same for the case of ${\lambda _A} > {\lambda _B}$. When we increase the reward of mint coins contained in key blocks (i.e. we increase $R$),  we observe that the optimal mining-duration lengths of the two pools approach $T=10$ for many parts of the mining power profile. This is because that large $R$ encourages mining in advance to earn the reward of mint coins in key blocks.

\begin{figure}[!t]
	\centering
	\includegraphics[width=3.2in]{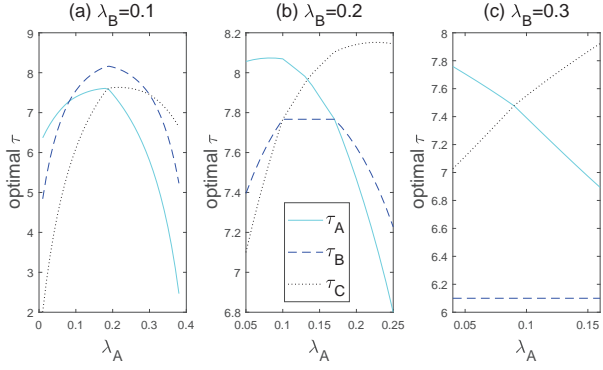}
	\caption{The optimal mining-duration lengths of the three mining pools are given as the functions of the mining power ${\lambda _A}$ for  ${\lambda _B} = \left\{ {0.1,0.2,0.3} \right\}$.}
	\label{fig_num3}
\end{figure}

We finally numerically compute the optimal mining-duration lengths for the $N$-player mining game by setting $N = 3$ (i.e., we have three mining pools denoted by $A$, $B$ and $C$). When we do that, we fix the mining power of the mining pool $B$ ${\lambda _B}$, and vary the mining power of mining pool $A$ ${\lambda _A}$. For a given ${\lambda _B}$, the computed optimal mining-duration lengths are functions of ${\lambda _A}$. The results are shown in Fig. \ref{fig_num3} when ${\lambda _B} = \left\{ {0.1,0.2,0.3} \right\}$ and $R=1$. We can see that for different power mining profiles, the mining pools need to employ different mining-duration lengths to achieve the equilibrium; mining pools with larger mining powers will converge to using shorter mining-duration lengths.

We conduct system simulations to investigate the advanced mining problem in the Bitcoin-NG network. Following the simulation approach used in \cite{eyal2018majority} for Bitcoin-like systems, we constructed a simulator that captures all the relevant Bitcoin-NG network details described in the previous sections, except that the crypto puzzle solving processing was replaced by a Monte Carlo simulator that simulates the time required for block discovery without actually attempting to compute a hash function. We simulated $2^{10}$ miners mining at identical rates (i.e., they each can have one simulated hash test at each time step of the Monte Carlo simulations and each test is a successful mining with probability $2^{-d}$). The total hash rate of the whole network thus is $2^{10}$. The difficulty level $d$ is set to ensure that in average one key block is found by the network during the mining time $T_m=2$, i.e., $\left( {{{{2^{10}}} \mathord{\left/
			{\vphantom {{{2^{10}}} {{2^d}}}} \right.
			\kern-\nulldelimiterspace} {{2^d}}}} \right){T_m} = 1$ which gives $d=11$.

\begin{figure}[!t]
	\centering
	\includegraphics[width=3.2in]{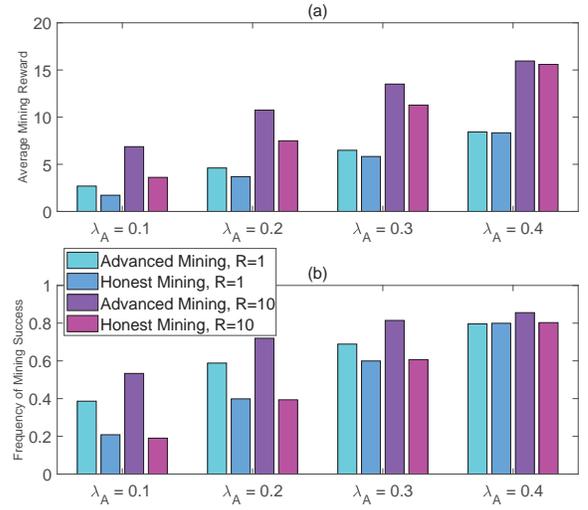}
	\caption{The simulation results for the one-attacker mining optimization problem: (a) the average mining reward of the attacker); (b) the frequency of mining success of the attacker.}
	\label{fig_sim1}
\end{figure}

The first simulation is to investigate the one-attacker mining optimization problem. In the simulation, the attacker's mining pool includes ${\omega _A}$ miners (thus its hash ratio is ${\lambda _A} = {{{\omega _A}} \mathord{\left/
		{\vphantom {{{\omega _A}} {{2^d} \in \left[ {0,0.5} \right]}}} \right.
		\kern-\nulldelimiterspace} {{2^d} \in \left[ {0,0.5} \right]}}$); the honest mining pool includes the remaining ${{2^{10}} - {\omega _A}}$ miners (thus its hash ratio is ${\lambda _B} = {{\left( {{2^{10}} - {\omega _A}} \right)} \mathord{\left/
		{\vphantom {{\left( {{2^{10}} - {\omega _A}} \right)} {{2^d}}}} \right.
		\kern-\nulldelimiterspace} {{2^d}}} = 0.5 - {\lambda _A}$). The reward contained in each key block is set to as $R=\{1,10\}$. The simulation results are given in Fig. \ref{fig_sim1} (a) for the average mining reward of the attacker and in Fig. \ref{fig_sim1} (b) for the frequency of mining success of the attacker. The hash ratio of the attacker is treated as a parameter and is set to as ${\lambda _A} = \{ 0.1,0.2,0.3,0.4\} $. The averaging is performed over $1000$ rounds. From Fig. \ref{fig_sim1} (a), we can see that for different hash ratios of the attacker, the attacker can get more profits by performing advanced mining than by performing honest mining; the gain of advanced mining over honest mining gets large, as the reward contain in key blocks $R$ gets large; the gain gets small, as the attacker's hash ratio ${\lambda _A}$ gets large.  From Fig. \ref{fig_sim1} (b), we can see that compared to having honest mining, the attacker increase its frequency of mining success by having advanced mining; the attacker with less computation power (smaller hash ratio) can increase more its frequency of mining success, compared to the attacker with higher computation power.

\begin{figure}[!t]
		\centering
		\includegraphics[width=3.1in]{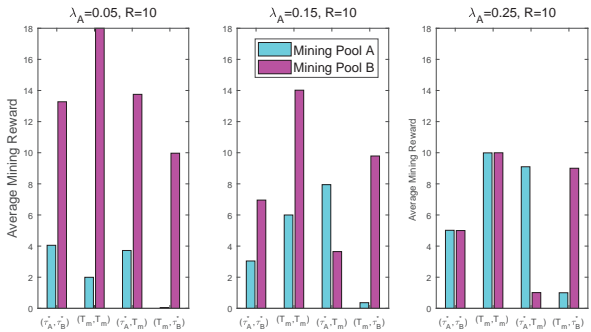}
		\caption{The simulation results of the average mining reward for the two-player mining game when $R = 10$.}
		\label{fig_sim2e}
	\end{figure}

\begin{figure}[!t]
		\centering
		\includegraphics[width=3.1in]{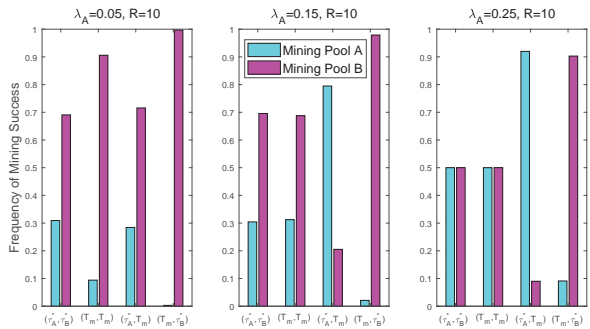}
		\caption{The simulation results of the frequency of mining success for the two-player mining game when $R = 10$.}
		\label{fig_sim2f}
	\end{figure}

The second simulation is to investigate the two-player mining game problem, where mining pools A and B compete with each other using advanced mining. Similarly, the mining pool A includes ${\omega _A}$ miners (thus its hash ratio is ${\lambda _A} = {{{\omega _A}} \mathord{\left/
		{\vphantom {{{\omega _A}} {{2^d} \in \left[ {0,0.5} \right]}}} \right.
		\kern-\nulldelimiterspace} {{2^d} \in \left[ {0,0.5} \right]}}$); the mining pool B includes the remaining ${{2^{10}} - {\omega _A}}$ miners (thus its hash ratio is ${\lambda _B} = {{\left( {{2^{10}} - {\omega _A}} \right)} \mathord{\left/
		{\vphantom {{\left( {{2^{10}} - {\omega _A}} \right)} {{2^d}}}} \right.
		\kern-\nulldelimiterspace} {{2^d}}} = 0.5 - {\lambda _A}$). We investigate the following four setups of mining strategies:
\begin{itemize}	
	
	\item[i)] $\left( {\tau _A^*,\tau _B^*} \right)$: both pools adopt advanced mining with their optimal mining-duration lengths given by the Nash equilibrium; 
	
    \item[ii)] $\left( {{T_m},{T_m}} \right)$: both pools adopt honest mining with the default mining-duration lengths; 
	
	\item[iii)] $\left( {\tau _A^*,{T_m}} \right)$: pool $A$ adopts advanced mining with its optimal mining-duration length given by the Nash equilibrium and pool $B$ adopts honest mining with default mining-duration length; 
	
	\item[iv)] $\left( {{T_m},\tau _B^*} \right)$: pool $A$ adopts honest mining with default mining-duration length and pool $B$ adopts advanced mining with its optimal mining-duration length given by the Nash equilibrium. 
	
\end{itemize}
The simulation results for the two mining pools under the above four setups of mining strategies are given in Fig. \ref{fig_sim2e}-\ref{fig_sim2f}. Under parameters ${\lambda _A}=\{0.05,0.15,0.25\}$ and $R=10$, the average mining rewards of the two mining pools are shown in Fig. \ref{fig_sim2e}, and the frequency of mining success of the two mining pools are shown in Fig. \ref{fig_sim2f}. We see from Fig. \ref{fig_sim2e} that the average mining rewards of the two mining pools with mining strategies $\left( {\tau _A^*,\tau _B^*} \right)$  are lower than that when  mining strategies $\left( {{T_m},{T_m}} \right)$ are adopted. If one pool deviates from advanced mining to honest mining, its average reward decreases when the other pool sticks to the advanced mining strategies. From the results in Fig. \ref{fig_sim2f}, we see that when the mining strategy of the other mining pool is fixed, the frequency of mining success of one particular mining pool is increased by its own advanced mining. Therefore, we conclude that both of the mining pools have no motivation to have honest mining due to their selfish and non-cooperative nature. From Fig. \ref{fig_sim2f}, we also find that a mining pool with less computation power (i.e., smaller hash ratio) has a higher motivation to perform advanced mining, since the advanced mining can increase its frequency of mining success regardless of what mining strategies the other mining pool adopts.

\begin{figure}[!t]
	\centering
	\includegraphics[width=3.1in]{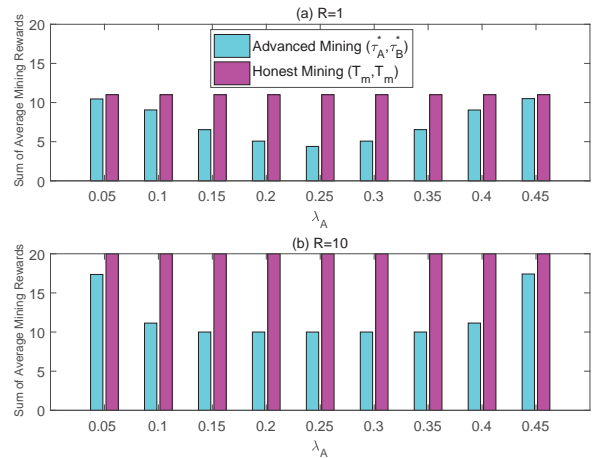}
	\caption{The simulation results of the sum average mining rewards of the two mining pools both adopting advanced mining and both adopting honest mining: (a) $R = 1$; (b) $R=10$.}
	\label{fig_sim222}
\end{figure}

Fig. \ref{fig_sim222} presents the sum and individual average mining rewards of mining pools $A$ and $B$ that both adopt advanced mining  ($\left( {\tau _A^*,\tau _B^*} \right)$) and or honest mining ( $\left( {{T_m},{T_m}} \right)$) for $R=1$ and $R=10$. For a particular setup of hash ratios, , e.g., $\lambda_A$ and $\lambda_B = 0.5 -\lambda_A $, there is a difference between the sum reward of the pools both having advanced mining $\left( {\tau _A^*,\tau _B^*} \right)$, and the sum reward of the pools both having honest mining $\left( {{T_m},{T_m}} \right)$. This difference in the sum rewards of the two mining pools is a system penalty caused by advanced mining. This penalty is in accordance with the reduced TPS---it is the number of the micro blocks that are not included in the blockchain. Comparing the average mining reward results in Fig. \ref{fig_sim222} (a) (for $R=1$) and in Fig. \ref{fig_sim222} (b) (for $R=10$), we can also see that the penalty caused by advanced mining increases as the increase of $R$.

Based on the observations, we can conclude: the advanced mining strategy setup $\left( {\tau _A^*,\tau _B^*} \right)$ is indeed the equilibrium of the mining game problem; advanced mining is harmful to the Bitcoin-NG network and it reduces the TPS of Bitcoin-NG; the TPS penalty caused by advanced mining decreases as the decrease of the reward contained in each key block ($R$). Therefore, if the total volume of the issued coins by Bitcoin-NG blockchain is fixed, releasing a small amount of new coins in each key block and letting the issue of coins last for a long time can alleviate the negative impact of advanced mining. Although zero key block reward can reduce the negative impact of advanced mining to the lowest, it will cause the blockchain unstable \cite{carlsten2016instability}.

\begin{figure}[t]
	\centering
	\includegraphics[width=3.1in]{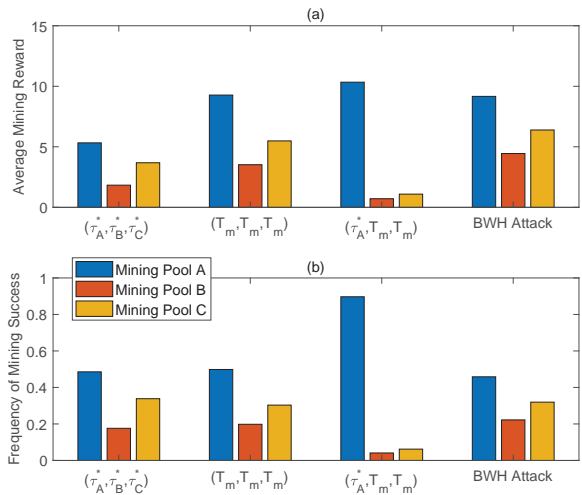}
	\caption{The simulation results for the three-player mining game when ${\lambda _A} = 0.25, {\lambda _B} = 0.1, {\lambda _C} = 0.15, R = 10$: (a) the average mining reward; (b) the frequency of mining success.}
	\label{three_player}
\end{figure}

We investigate the mining game of $N$-players by considering the scenario where three mining pools $A$, $B$ and $C$ (i.e. $N=3$) play the mining game. In the simulation, the three mining pools have hash ratios ${\lambda _A} = 0.25 $, ${\lambda _B} = 0.1$, and ${\lambda _C} = 0.15 $, respectively. We consider three setups of mining strategies for them: i) all three mining pools adopt advanced mining ($({\tau _A^*,\tau _B^*,\tau _C^*} )$);  ii) all three mining pools adopt honest mining ($(T_m,T_m,T_m )$); iii) mining pool $A$ adopts advanced mining and the other two adopt honest mining  ($(\tau _A^*,T_m,T_m )$), where $\tau _A^*$, $\tau _B^*$, and $\tau _C^*$ are the optimal mining-duration lengths given by the Nash equilibrium. The simulation results are given in Fig. \ref{three_player} when $R=10$, where Fig. \ref{three_player} (a) presents the results of the average mining reward, and Fig. \ref{three_player} (b) presents the results of the frequency of mining success. From the simulation results, we can see that for the mining game where $N=3$, the best strategy of a particular player (e.g., mining pool $A$) is advanced mining. This validates the Nash equilibrium for the mining game where $N=3$.

We also compare advanced mining attack with BHW attack [22] in the three-player scenario. For the game-theoretical model of BHW attack, we assume that the mining pools $B$ and $C$ can access the mining pool $A$ to adopt the optimal power-splitting strategies proposed in [22] to play the mining game. The simulation results of the BHW attack game are presented in Fig. 10. We can observe that in the simulated three-player scenario, the BHW attack game gives less severe attack effects than the advanced mining game dose, i.e, the advanced mining game changes the distribution of mining rewards more severely.

\section{Conclusion}

In this work, we investigated the advanced mining problem for the Bitcoin-NG blockchain protocol. Although Bitcoin-NG is scalable, it is vulnerable to the malicious advanced mining in which attackers intentionally ignore some micro blocks issued by the current leader and mine on a early micro block to enlarge their successful mining probabilities. We find that although advanced mining will lose some transaction fees contained in later micro blocks, it is still more profit than honest mining (i.e., mining on the latest micro block). Moreover, we show that when mining pools adopt advanced mining, the mining problem of Bitcoin-NG can be formulated as a non-cooperative game and each mining pool individually decides when to mine. We then find the equilibrium of this mining game. Numerical results are provided to confirm our analytical results. We have performed system simulations to investigate the advanced mining problem. Based on the simulation results, we have the following conclusions: the analytically derived advanced mining strategy is indeed the equilibrium of the mining game; advanced mining will reduce the TPS of Bitcoin-NG; mining pools with less computation power have higher motivations to perform advanced mining.

\ifCLASSOPTIONcaptionsoff
  \newpage
\fi

\bibliographystyle{IEEEtran}

\bibliography{refs}

% that's all folks
\end{document}